# Low-filling-factor superconducting single photon detector with high system detection efficiency

Taro Yamashita,[1,*] Shigehito Miki,[1] Hirotaka Terai,[1] and Zhen Wang[1,2]

[1]Advanced ICT Research Institute, National Institute of Information and Communications Technology, 588-2 Iwaoka, Nishi-ku, Kobe 651-2492, Japan
[2]Shanghai Institute of Microsystem and Information Technology, Chinese Academy of Sciences, 865 Changning Road, Shanghai 200050, China
[*]taro@nict.go.jp

**Abstract:** We designed, fabricated, and measured superconducting nanowire single-photon detectors (SSPDs) with low filling factor which achieve high system detection efficiency (SDE) and counting rate simultaneously. Numerical simulation reveals that high optical absorptance is possible in SSPDs even for low filing factor by tuning the device design. The SDEs of fabricated 18-50% filling factor SSPDs were measured systematically, and all SSPDs showed high SDEs of 61-80% and the lowest 18% filling factor SSPD achieved a high SDE of 69%.

## 1. Introduction

In various research fields, such as quantum communication [1], quantum optics [2], life science [3] etc., superconducting nanowire single-photon detectors (SSPDs or SNSPDs) are actively used because of their high system detection efficiency (SDE), low dark count, high counting rate (speed), and excellent timing resolution [4,5]. Among these features, the primarily important performance to be improved is SDE, defined as SDE = $P_{pulse} \times P_{couple} \times P_{abs} \times (1 - P_{loss})$, where $P_{pulse}$ is the pulse generation probability of the nanowire, $P_{couple}$ the coupling efficiency between the incident light and the active area, $P_{abs}$ the optical absorptance in the nanowire, and $P_{loss}$ any other optical losses in the system. Therefore, improvements of all factors are necessary to achieve a high SDE. Regarding $P_{abs}$, a key point for achieving a high optical absorptance is the configuration of the optical cavity structure in the device [6,7]. Recently, a double-side cavity structure was adopted to obtain a high $P_{abs}$, resulting in excellent SDEs of 93% [8], 76% [9], and 74% [10]. In the double-side cavity SSPDs, the superconducting nanowire layer was sandwiched between upper and lower cavity layers of dielectric materials, such as SiO and $SiO_2$. The incoming photons were strongly confined in the optical cavity; hence, a large $P_{abs}$ can be achieved. The filling factor of the meandering nanowire, which is defined as the line width divided by the pitch of the line and their spacing, is also important for achieving a high $P_{abs}$. So far, the relatively high filling factors have been adopted for SSPDs to achieve the high optical absorptance by the following reason. From a physical point of view, $P_{abs}$ is proportional to the integral of the square of the electric field in the nanowire region [11]. When the filling factor is reduced, the proportion of the nanowire in the whole active area becomes smaller. For the incident light with the fixed power density, the electric field is concentrated further around the nanowire by reducing the filling factor, and thus the average electric-field intensity in the nanowire increases as discussed precisely in the next section. However, if the degree of the electric-field enhancement is small, eventually the optical absorptance decreases because the relative nanowire region becomes smaller with reduced filling factor [11]. Thus, conventional SSPDs have relatively large filling factors of around fifty percent or higher to obtain high SDEs [8-10] though the electrical properties have been investigated for low-filling-factor device [12].

Another important performance index in SSPDs is the counting rate. In conventional SSPDs, the counting rate is limited by a large kinetic inductance due to the long meandering nanowire to realize the large active area [13]. The large active area in SSPD is required to achieve a high $P_{couple}$, and hence a high SDE. Actually, SSPDs for use in practical application have a millimeter-long nanowire [8,10] and their maximum counting rate is around several ten megahertz, which is far from the gigahertz expected intrinsically. Therefore, there has been a trade-off problem to achieve simultaneously a high counting rate and a high SDE. In order to achieve the high counting rate without degrading the SDE, several device structures have been proposed, i.e., multi-element array to reduce the active area of each detector element [9], parallel nanowire configuration [14], optical nano-antenna to keep the good $P_{abs}$ for the low-filling-factor layout [15], and the short nanowire optically-coupled to the waveguide [16,17]. However, their structures become complex and/or performances are still developing compared to the conventional SSPDs. In this paper, we present a strategy to solve the trade-off between SDE and the counting rate by using a strongly enhanced electric field in the double-side cavity structure. We show how the high optical absorptance can be achieved for the low-filling-factor layout by the numerical simulation, and fabricate the SSPDs with various filling factor, and then evaluate and analyze their performances systematically.

## 2. Numerical simulation of optical absorptance

In order to investigate the relation between the optical absorptance and the filling factor, we calculated the electric field and optical absorptance in the nanowire numerically by using finite element analysis. The numerical simulation has been employed in several works to explore the effective device structures to achieve the high optical absorptance [11,18-21]. In the present work, we used the COMSOL Multiphysics 4.3a with RF module reported in Refs. 11 and 21 as the numerical simulation software. Figure 1(a) shows the simulated structure corresponding to SSPDs consisting of a double-side cavity structure with back-side illumination [10]. In the numerical simulation, we assumed a two-dimensional unit cell consisting of the Si substrate, $SiO_2$, NbN nanowire, SiO, and Ag mirror from bottom (the illumination side) to top (the region surrounded by the dashed line in Fig. 1(a)). The cavity $SiO_2$ and SiO layers were 240-nm and 230-nm thick, respectively. They were chosen to enhance the optical absorptance in the nanowire at a 1.55-µm wavelength with a 130-nm-thick Ag mirror. The line width of the NbN nanowire was fixed to 80 nm and the periodic boundary condition was used for the boundaries of the unit cell. The incident light was assumed to be polarized parallel to the nanowire and to enter the device from the Si substrate.

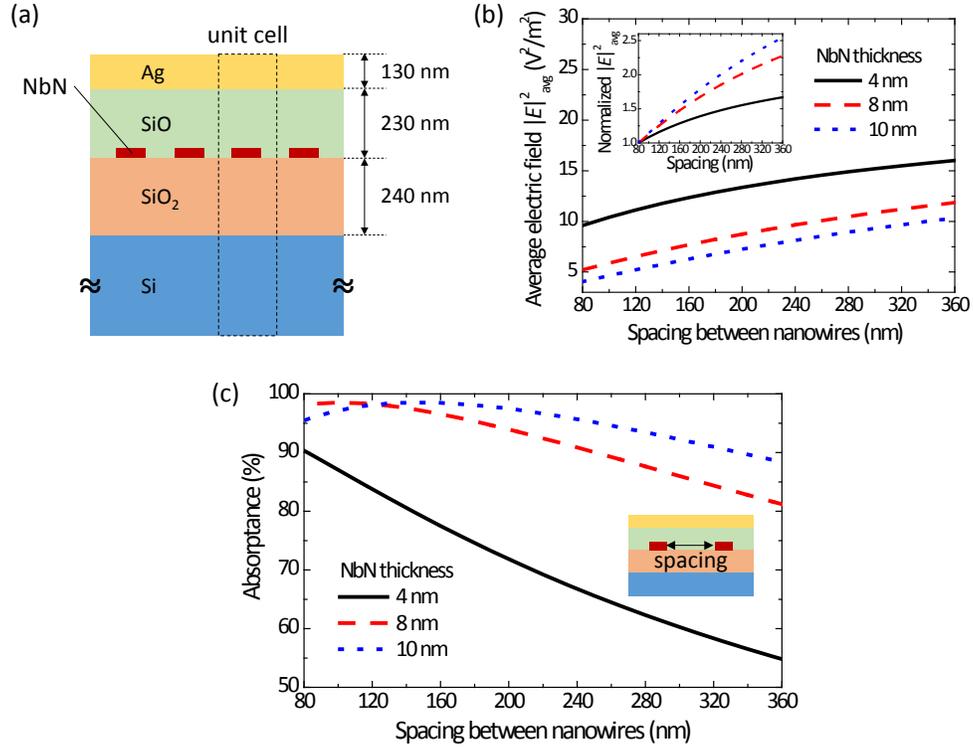

Fig. 1. (a) Schematic of the stack structure of a SSPD with a double-side cavity. The region surrounded by the dashed line indicates the unit cell for the numerical simulation. The NbN nanowire width was fixed to 80 nm. The thickness of the $SiO_2$, SiO, and Ag-mirror layers were set to 240 nm, 230 nm, and 130 nm, respectively. In the simulation, we used complex refractive indices $n_{Si} = 3.628$, $n_{SiO_2} = 1.444$, $n_{SiO} = 1.551$, $n_{NbN} = 4.905 + i4.293$, and $n_{Ag} = 0.322 + i10.99$. The indices were measured by spectroscopic ellipsometry. The incident light, polarized parallel to the nanowire, was assumed to enter the device from the Si substrate. (b) The spacing dependences of the spatial average of the square of the electric field, $|E|^2_{avg}$. The solid black, dashed red, and dotted blue curves are for the thicknesses of 4 nm, 8 nm, and 10 nm, respectively. Inset: the spacing dependences of $|E|^2_{avg}$ normalized by that for 80-nm spacing. (c) Simulated spacing dependences of the optical absorptance in the NbN nanowire. The solid black, dashed red, and dotted blue lines represent the 4-nm-, 8-nm-, and 10-nm-thick NbN nanowires, respectively.

First we investigated the electric field generated in the NbN nanowire. Assuming the electric field in the small nanowire region is almost uniform, the optical absorptance is approximately proportional to the product of the filling factor and the spatial average of the square of the electric field, $|E|^2_{avg}$ [11]. Figure 1(b) shows the spacing (pitch minus line width of 80 nm) dependences of $|E|^2_{avg}$ for the NbN thicknesses of 4 nm, 8 nm, and 10 nm. As shown in the figure, the electric field generated in the nanowire is enhanced for increasing the spacing, i.e., reducing the filling factor because the electric field is concentrated into the nanowire further when the occupancy of the nanowire region in the whole device (unit cell) region becomes smaller. Similarly, $|E|^2_{avg}$ increases for decreasing the thickness for the fixed filling factor. In order to investigate the degree of the enhancement of $|E|^2_{avg}$ by reducing the filling factor, we plotted the pitch dependences of $|E|^2_{avg}$ normalized by that for 80-nm spacing (50% filling factor) in the inset of Fig. 1(b). As seen from the inset, the degree of the electric-field enhancement by reducing the filling factor is larger for the thicker nanowire. For example, for the 4-nm thick nanowire which is typical for conventional SSPDs, when the filling factor is reduced to 25% (240-nm spacing) from 50%, $|E|^2_{avg}$ is increased by less than 1.5 times. On the other hand, for the thicker nanowires, the degree of the enhancement is larger, e.g., in the 10-nm-thick nanowire $|E|^2_{avg}$ becomes twice by reducing the filling factor to 25% from 50%.

Figure 1(c) shows a simulation of how the optical absorptance in the NbN nanowire depends on the spacing between the nanowires (pitch minus line width of 80 nm) for 1.55-μm wavelength incident light. We calculated the absorptance for the nanowire with thickness of 4 nm (solid black line). The optical absorptance showed a highest value of 90% at the shortest space of 80 nm corresponding to a filling factor of 50%. When the spacing between the nanowires increased, i.e., when the filling factor is reduced, the optical absorptance decreased monotonically. The absorptance dropped to 55% for the nanowires with a large spacing of 360 nm. On the other hand, the optical absorptance showed a different dependence on the filling factor by increasing the thickness of the nanowire. The dashed red line in Fig. 1(c) indicates the absorptance for an 8-nm-thick nanowire. In this case, the optical absorptance shows a more gradual dependence on the spacing between nanowires, and the absorptance was kept at high values of over 80% even in the lower filling factor region. Even at the 360-nm spacing corresponding to a very low filling factor of 18%, the optical absorptance reached 81% which is much higher than for the 4-nm-thick nanowire. Furthermore, for the nanowire with 10-nm thickness (dotted blue line), the optical absorptance did not decrease monotonically with increasing spacing and showed a peak around a spacing of 145 nm (filling factor of 36%). The absorptance for the 360-nm spacing is improved further to ~ 90%, which is not so far from the maximum absorptance of 98% for the 145-nm spacing. The improvement of the optical absorptance by increasing the nanowire thickness can be explained by the competition between the filling factor and the electric-field enhancement; when the filling factor is reduced, the electric field generated in the nanowire region is enhanced as mentioned before. The degree of the enhancement depends on the thickness of the nanowire, and becomes larger with increasing the thickness. Thus, the thick nanowires with 8-nm and 10-nm thickness show a more gradual absorptance dependence on the filling factor whereas the absorptance in the 4-nm-thick nanowire decreases more rapidly with reduced filling factor. Especially for the 10-nm-thick nanowire, the degree of the electric-field enhancement exceeds the effect of the reduction of the filling factor for spacings from 80 nm to 145 nm, and a non-monotonic dependence with a peak structure appears.

## 3. Experimental setup

Based on the simulation result of the optical absorptance, we fabricated 8.5-nm-thick NbN SSPDs on Si substrates with thermally oxidized silicon ($SiO_2$) layers for both surfaces. The $SiO_2$ layer at the back (illuminated) side worked as an anti-reflection layer and that at the front side as a cavity layer. The NbN films were deposited by DC reactive sputtering at an ambient substrate temperature. The thickness of the deposited NbN films was controlled by the deposition rate and time. The NbN films were formed into a meandering nanowire with 80-nm width covering a square area of 15 μm × 15 μm. In order to verify the validity of the theoretical prediction for the dependence of the optical absorptance on the spacing between the nanowires, we prepared the devices with various spacings of 80, 120, 160, 200, 240, 280, 320, and 360 nm (filling factors of 18 - 50%). The superconducting transition temperatures, $T_c$, were 7.0 – 7.2 K and the switching currents were 11.4 – 14.0 μA (the current densities of 1.7 – 2.1 × $10^{10}$ A/m$^2$) for the fabricated devices. In order to check the actual scale of the fabricated device, we observed a 240-nm-spacing SSPD by transmission electron microscope (TEM) and confirmed that the device was fabricated in accordance with the intended design.

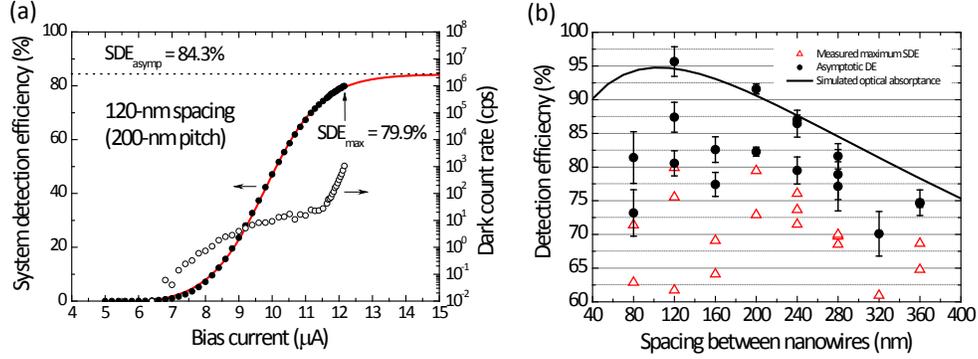

Fig. 2. (a) System detection efficiency (SDE) and dark count rate vs. bias current for the SSPD with 120 nm spacing. The measured maximum SDE ($SDE_{max}$) was 79.9% for the dark count rate of ~ 1 kcps and the asymptotic SDE ($SDE_{asymp}$) evaluated from fitting by a sigmoid function was 84.3%. (b) Dependences of $SDE_{max}$ (red open triangles) and asymptotic DE (black filled circles) on the spacing between the nanowires. The error bars for the asymptotic DE indicate the uncertainty of the measured optical losses at low temperatures. The solid line is the numerically simulated optical absorptance including the assumed 4% coupling loss.

To evaluate the performances of the fabricated SSPDs, the device chip was mounted in fiber-coupled compact packages using fiber-spliced graded index (GRIN) lenses to obtain good optical coupling efficiency between the illuminated light and the device active area, $P_{couple}$ [22]. The packaged devices were then installed into a 0.1-Watt Gifford-McMahon cryocooler system. In this system, the sample stage was cooled down to 2.1 K within a thermal fluctuation range of 10 mK. Semi-rigid coaxial cables and single-mode optical fibers were introduced to each package, and the optical fibers in the cryocooler system made several loops, with a diameter of around 30 mm, to mitigate dark counts due to blackbody radiation. The semi-rigid coaxial cables were connected to a bias tee and two room-temperature low noise amplifiers outside the cryocooler. When measuring the SDE, a continuous tunable laser was used as photon source with its light power heavily attenuated so that the photon flux at the input connector of the cryostat was $10^6$ photons/s. A polarization controller was put in front of the cryocooler optical input to control the polarization of the incident photons to maximize the SDE. Although it is difficult to measure an actual polarization at the end of the fiber, we expect that the polarization is parallel to the nanowire (TE) where the absorptance is maximum from the numerical simulation. The SSPD output counts were measured by a pulse counter. In order to avoid the influence of periodic fringes, resulting from the interference among the multi optical layer boundaries, the wavelength was tuned to maximize the SDE in the range from 1540 to 1560 nm [7]. The SDE was evaluated by the relation SDE = ($R_{output}$ − $R_{DCR}$)/$R_{input}$, where $R_{output}$ is the SSPD output pulse rate, $R_{DCR}$ the dark count rate, and $R_{input}$ the input photon rate to the system.

## 4. Results and discussion

*4.1 Detection efficiency*

Figure 2(a) shows the measured bias-current dependences of the SDE (black symbols) and the dark count rate (open circles) for SSPD with spacing of 120 nm (200-nm pitch). The measured maximum SDE ($SDE_{max}$) reached a high value of 79.9% at the dark count rate of ~ 1 kcps near the switching current [23]. In this device, the SDE curve does not saturate even at near the switching current, indicating that the intrinsic pulse generation probability $P_{pulse}$ of this device does not attain 100%. To estimate the potentially maximum SDE when $P_{pulse} = 1$

(asymptotic SDE; $SDE_{asymp}$), we fitted the SDE data by a sigmoid function (red line in Fig. 2(a)) [10]. $SDE_{asymp}$ is defined as SDE at the saturated region in the fitted sigmoid function, and we obtained $SDE_{asymp}$ of 84.3% by the fitting. The sigmoidal type of shape for the bias-current dependence of SDE has been known empirically [8-10]. Although the physical basis for this function is not clear, only $P_{pulse}$ depends on the bias current in SDE, and thus we could expect that $P_{pulse} = 1$ in the saturated region because the detection efficiency is not improved even by applying the bias current further. From the relation $P_{pulse} = SDE_{max}/SDE_{asymp}$, we obtained the high $P_{pulse}$ of 94.8% for this device. So far, there was a concern that $P_{pulse}$ in the thick nanowire (e.g., 8.5 nm in the present work) may be small and limit the SDE because the superconductivity of the large cross-section area nanowire is robust for the energy excitation by the photons. However, even for the thick nanowires, a good $P_{pulse}$ is expected by making $T_c$ relatively low, and actually all developed 8.5-nm-thick SSPDs with $T_c$ of ~ 7 K showed a high $P_{pulse}$ of 91 - 98%. A thick SSPD with a high $P_{pulse}$ is expected also by using low-$T_c$ materials such as WSi [8] or by tuning the deposition condition to make the superconductivity in the nanowire weaker. Reducing the nanowire width further is also a potential way to achieve the high $P_{pulse}$ for thick nanowires [21,24].

As mentioned before, the SDE is expressed as $SDE = P_{pulse} \times P_{couple} \times P_{abs} \times (1 - P_{loss})$, and therefore $SDE_{asymp}$ is equal to $P_{couple} \times P_{abs} \times (1 - P_{loss})$. Here we can assume $P_{couple} \approx 1$ which leads to $SDE_{asymp} = P_{abs} \times (1 - P_{loss})$ since the focused light spot size of 9-μm diameter is smaller than the device active area of 15 μm × 15 μm and thus all incident light is expected to enter the active area [22]. In order to compare the obtained experimental data with the simulation results of the optical absorptance, we define $DE_{asymp} = SDE_{asymp}/(1 - P_{loss})$ in which the optical losses in the system are excluded from $SDE_{asymp}$. We measured $P_{loss}$ at a low temperature for each optical port and obtained losses of 0.1 - 0.7 dB in the system. The $DE_{asymp}$ is more appropriate for comparing with simulated results than the measured $SDE_{max}$, which includes the chip-by-chip variation in $P_{pulse}$ possibly due to variations in the degree of the constriction as well as the optical losses.

Figure 2(b) shows the dependences of the detection efficiencies on the spacing between the nanowires. The open red triangles and filled black circles indicate the measured $SDE_{max}$ and derived $DE_{asymp}$, respectively. Note that the data plotted in Fig. 2(b) is for all devices we measured, indicating the yield of the device is quite good. As shown in the figure, the highest $SDE_{max}$ of 79.9% was obtained for the 120-nm-spacing SSPD, and both $SDE_{max}$ and $DE_{asymp}$ showed similar gradual dependences on the spacing between the nanowires. Surprisingly, the $SDE_{max}$ of the 360-nm-spacing device with the lowest filling factor of 18% was a high value of 68.7%, which is only 11% lower than the highest $SDE_{max}$ for the 120-nm-spacing device. Furthermore, the $DE_{asymp}$ of this low-filling-factor device reached as high as 74.7%. In the future, we expect $SDE_{max}$ to be improved further by removing the currently used fiber connectors causing an optical loss, and using a spliced fiber connecting directly to the SSPD from the outside of a cryocooler. In order to analyze the experimental result, we performed a numerical simulation of the optical absorptance. As shown in the solid black line in Fig. 2(b), the simulated curve followed the experimental spacing dependence of $DE_{asymp}$, with the assumed 4% coupling loss as only one fitting parameter [8]. From the above experimental results and theoretical analysis, it was confirmed that the 8.5-nm-thick and low-filling-factor SSPD achieved an excellent SDE because of the strong enhancement of the electric field by the double-side cavity structure.

*4.2 Counting rate*

Next we measured the kinetic inductance of SSPDs expressed as $L_k = \mu_0 \lambda^2 l/S$, where $\mu_0$ is the vacuum permeability, $\lambda$ is the penetration depth, $l$ is the nanowire length, and $S$ is the cross-section area of the nanowire. $L_k$ can be estimated by measuring the phase of the reflection coefficient $S_{11} = (i\omega L_k - 50\Omega)/(i\omega L_k + 50\Omega)$ by using a network analyzer [25]. By fitting the expression of $S_{11}$ to the experimental data with $L_k$ as a free parameter, $L_k$ can be derived. Figure 3(a) indicates the dependence of the evaluated $L_k$ on the nanowire length $l$. As expected from the expression of $L_k$, the obtained values of $L_k$ are proportional to the nanowire length from the origin. The devices with the shortest, 511-μm, nanowire (i.e., lowest filling factor of 18%) have the smallest $L_k$ of 0.71 μH, which is almost one-third of the $L_k$ of 2.0 μH for the 50%-filling- factor SSPDs with a 1406-μm long nanowire.

Finally, to estimate the counting rate, we measured the dependences of the SSPD counts on the incident photon rate [9]. In this measurement, we used the continuous laser as a photon source and controlled the incident photon rate by adjusting the attenuator, and then the SSPD counts were measured by a pulse counter. Figure 3(b) shows that the measured SSPD counts vs. incident photon rate for the highest (50%) and lowest (18%) filling factor SSPDs. Both SSPDs were biased to show the SDE of 50%. For the small photon rate, the measured counts increase in proportion to the incident photon rate since the incoming photons enter the SSPD with enough time intervals for the SSPD to recover its superconducting state. When the incident photon rate increases, it becomes difficult for the SSPD to respond to all incoming photons correctly because after detecting the first photon a second one can enter the device within the recovery (dead) time of the SSPD. As shown in Fig. 3(b), for the 50%-filling-factor SSPD with the long nanowire (1406 μm, black line), the curve deviated from the proportional relation for incident photon rates over 10 MHz. On the other hand, it is clearly shown that the curve for the 18%-filling-factor SSPD with the short nanowire (511 μm, red line) kept the proportional relation for the higher photon rate, indicating that the low-filling-factor SSPD achieved the improved counting rate.

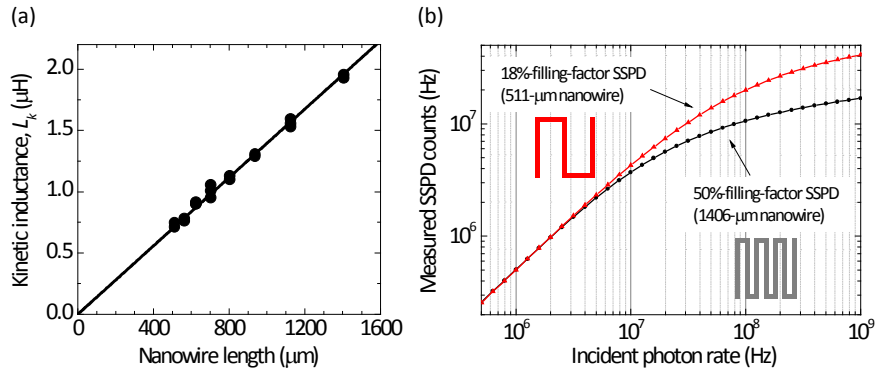

Fig. 3. (a) Measured kinetic inductance vs. nanowire length (symbols). The solid line indicates fitting curve by the expression for $L_k$. (b) Incident photon rate dependences on the SSPD output counts. The black and red curves indicate the SSPD counts for the 50%-filling-factor device with a 1406-μm nanowire and the 18%-filling-factor device with a 511-μm nanowire, respectively.

## 5. Conclusions

In conclusion, we have presented a low-filling-factor SSPD, which can achieve an excellent SDE and high counting rate in the same device. By the numerical simulation of the optical absorptance, we found that a high absorptance can be achieved even for the low-filling-factor SSPDs by tuning the device design. Based on a simulation result, we fabricated double-side cavity SSPDs with 8.5-nm-thick NbN nanowire, which is thicker than the conventional 4 - 5 nm SSPDs. For SSPDs with different spacing between the nanowires, the detection efficiencies were explained well by the simulation results and showed an excellent SDE of 68.7% for the lowest filling factor of 18%. The low-filling-factor design is effective not only for improving the counting rate, but it is also expected to reduce difficulties in fabricating an uniform nanowire since it becomes harder to avoid the defects in longer nanowires. Our result presented in this paper provides a breakthrough to achieve high-speed and excellent-SDE single photon detectors useful in a wide range of applications.

**Acknowledgments**

The authors thank Saburo Imamura and Makoto Soutome for technical support and Dr. Satoshi Ishii for helpful discussions on the numerical simulation.